\documentstyle[aps,prl,epsf,floats]{revtex}

\begin{document}
\draft

\twocolumn[\hsize\textwidth\columnwidth\hsize\csname
@twocolumnfalse\endcsname

\title{ {\small\hspace{14.5cm} BIHEP-EP1-2000-01} \\
\bf\boldmath A Measurement of the Mass and Full-Width of the $\eta_c$ Meson}

\author{
J.~Z.~Bai,$^1$      Y.~Ban,$^{11}$      J.~G.~Bian,$^1$
I.~Blum,$^{19}$
A.~D.~Chen,$^1$     G.~P.~Chen,$^1$     H.~F.~Chen,$^{18}$  H.~S.~Chen,$^{1}$
J.~Chen,$^5$
J.~C.~Chen,$^1$
% ...
\\(BES Collaboration)\\ }

\address{
$^1$ Institute of High Energy Physics, Beijing 100039, People's Republic of
     China\\
% ...
$^5$ Colorado State University, Fort Collins, Colorado 80523\\
% ...
$^{11}$ Peking University, Beijing 100871, People's Republic of China\\
% ...
$^{18}$ University of Science and Technology of China, Hefei 230026,
        People's Republic of China\\
$^{19}$ University of Texas at Dallas, Richardson, Texas 75083-0688\\
% ...
}


\maketitle

\begin{abstract}
In a sample of 7.8 million $J/\psi$ decays collected in
the Beijing Spectrometer, the process
J/$\psi\to\gamma\eta_c$ is
observed for five different $\eta_c$  decay channels:
$K^+K^-\pi^+\pi^-$, $\pi^+\pi^-\pi^+\pi^-$, 
$K^\pm K^0_S \pi^\mp$ (with $K^0_S\to\pi^+\pi^-$),
$\phi\phi$ (with $\phi\to K^+K^-$) and
$K^+K^-\pi^0$.  From these signals,
we determine the mass of $\eta_c$ to be $2976.6\pm2.9\pm1.3$ MeV.
Combining this result with a previously reported
result from a similar study using $\psi(2S)\to\gamma\eta_c$ 
detected in the same spectrometer gives
$m_{\eta_c} = 2976.3\pm2.3\pm1.2$ MeV.  For the combined samples,
we obtain
$\Gamma_{\eta_c} = 11.0\pm 8.1\pm 4.1$ MeV. 
\end{abstract}

]

A precise knowledge of the mass difference between
the J/$\psi(1^{--})$  and $\eta_c(0^{-+})$ charmonium states
is useful
for the determination of the strength of the spin-spin interaction term
in non-relativistic potential models.  While the
$J/\psi$ mass has been determined with high accuracy (1 part in $10^{5}$) to be
$3096.88 \pm 0.04$ MeV,  the mass of the $\eta_c$ is less well measured. The
Particle Data Group
(PDG) average of $m_{\eta_c} =2979.8 \pm 2.1$ MeV is based on
experiments using the reactions $e^+e^-\to$
J/$\psi\to\gamma\eta_c$ \cite{dm2,m3-1,m3-2,m3-3,cball},
$e^+e^-\to\psi(2S)\to\gamma\eta_c$ \cite{cball}
and p$\bar{\mbox p}\to\gamma\gamma$ \cite{e760-1,e760-2}. 
These measurements have poor internal consistency, and the PDG fit to
the measurements has a confidence level of only 0.001. The most recent
result from Fermilab experiment E760 \cite{e760-1} 
disagrees with the result from the DM2 group \cite{dm2}
by almost four standard deviations.  Measurements of the
full width of the $\eta_c$ have been made by four groups:
E760 reports a result of
$\Gamma_{\eta_c} = 23.9_{-7.1}^{+12.6}$~MeV \cite{e760-1},
which is larger than the results from
SPEC ($7.0_{-7.0}^{+7.5}$~MeV) \cite{e760-2}, 
Mark III ($10.1_{-8.2}^{+33.0}$~MeV) \cite{m3-1} and Crystal Ball 
($11.5\pm 4.5$~MeV)\cite{cball}.
Additional measurements for both $m_{\eta_c}$
and $\Gamma_{\eta_c}$ are needed to improve the situation.  An
$\eta_c$ mass value of $m_{\eta_c} = 2975.8 \pm 3.9 \pm 1.2$ MeV
was reported earlier by the Beijing Spectrometer (BES)
collaboration based on an
analysis of the reaction $\psi$(2S)$\to\gamma\eta_c$ \cite{bespsi}. 
In this paper we report a measurement of the mass of the
$\eta_c$ based on a data sample of 7.8 million J/$\psi$ events collected in BES.
The reactions
J/$\psi\to\gamma\eta_c$, $\eta_c\to K^+K^-\pi^+\pi^-$, $\pi^+\pi^-\pi^+\pi^-$,
$K^\pm K^0_S \pi^\mp$ (with $K^0_S\to\pi^+\pi^-$), 
$\phi\phi$ (with $\phi\to K^+K^-$)  and $K^+K^-\pi^0$ 
have been used to determine the mass and width of the $\eta_c$.

The Beijing Spectrometer  has been described
in detail in Ref.~\cite{detect}.  Here we describe briefly those detector
elements essential to this measurement. 
Charged particle tracking is provided by a 10 superlayer
main drift chamber (MDC). Each superlayer contains four 
cylindrical layers of sense wires that
measure both the position and the ionization energy loss (dE/d$x$) of
charged particles. The momentum resolution is
$\sigma_P/P = 1.7\%\sqrt{1 + P^2}$, where $P$ is in GeV/$c$. 
The dE/d$x$ resolution is 
$9\%$ and provides good $\pi/K$ separation in the
low momentum region. An array of 48 scintillation counters surrounding the
MDC measures the time-of-flight (TOF) of charged tracks with a resolution of
330 ps for hadrons. Outside of the TOF system is an electromagnetic calorimeter
comprised of streamer tubes and lead sheets with a $z$ position resolution
of 4 cm. The energy resolution of the shower counter scales as $\sigma_E/E
= 22\%/\sqrt{E}$, where
$E$ is in GeV. Outside the shower counter is a solenoidal magnet that
produces a 0.4 Tesla magnetic field.

The event selection criteria for each channel are described in detail in previous
papers \cite{dly2kp,dly2pp,dlyk3p}.
Here we repeat only the essential information and emphasize those  
considerations that are special to the $m_{\eta_c}$ measurement.
Candidates are selected by requiring the correct number of
charged track candidates for the given hypothesis.  These
tracks must be well fit to a helix in the polar angle
range $-0.8 < \cos\theta < 0.8$ and have a transverse momentum 
above $60$ MeV/c.
For the  four-charged-track channels, at least one photon 
with energy $E_{\gamma}> 30$~MeV is required in the 
barrel shower counter; for the $K^+K^-\pi^0$ channel, at least three
$E_{\gamma}>30$ MeV photons
are  required.
Showers that can be associated with charged tracks are not considered.
Events are fitted kinematically with four constraints (4C) to the hypotheses:
$J/\psi \to \gamma K^+K^-\pi^+\pi^-$, 
$J/\psi \to \gamma \pi^+\pi^-\pi^+\pi^-$, 
$J/\psi \to \gamma K^\pm\pi^\mp\pi^+\pi^-$,
$J/\psi \to \gamma \gamma \gamma K^+K^-$.
A one-constraint (1C) fit is done for the 
$J/\psi \to \gamma_{miss} K^+K^-K^+K^-$ hypothesis,
where $\gamma_{miss}$ indicates that this photon is not detected.
We select those events for each particular channel 
that have a confidence level greater than 5\%.
A cut on the variable, $|U_{miss}|=|E_{miss}-P_{miss}| < 0.10$ 
GeV (for $\pi^+\pi^-\pi^+\pi^-$),
$<0.12$ GeV (for $K^+K^-\pi^+\pi^-$), 
$<0.15$ GeV (for $K^\pm K_{S}^{0}\pi^\mp$) and
$<0.15$ GeV (for $\phi\phi$) is imposed 
to reject events with multiphotons and misidentified charged particles.
Here, $E_{miss}$ and $P_{miss}$ are, respectively, the missing energy and 
missing momentum calculated using the measured quantities
for the charged tracks.
Another cut on the variable,
$P_{t\gamma}^2 = 4|P_{miss}|^2\sin^2(\theta_{t\gamma}/2) < 0.006 (\mbox{GeV}/c)^2$
(for $K^+K^-\pi^+\pi^-$, $\pi^+\pi^-\pi^+\pi^-$ and $K^\pm K_{S}^{0}\pi^\mp$)
is used to reduce the backgrounds from $\pi^0$'s, where $\theta_{t\gamma}$
is the angle between the missing momentum and the photon direction.
For the $K^+K^-\pi^+\pi^-$  and $\pi^+\pi^-\pi^+\pi^-$ channels,
$|M_{\pi^+\pi^-\pi^0}-M_\omega|>30$ MeV is required to remove the background
from $J/\psi\to \omega \pi^+\pi^-$ and $J/\psi\to \omega K^+K^-$;
where a $\pi^0$ is associated with the missing momentum.
For the $K^\pm K_{S}^{0}\pi^\mp$ (with $K^0_S\to\pi^+\pi^-$) channel, the  
$\pi^+\pi^-$ invariant mass for the
$K^0_S $ candidate is required to be within 25~MeV
of $M_{K_{S}^{0}}$.
For the $\phi\phi$ (with $\phi\to K^+K^-$) channel, 
the invariant masses of both candidate $\phi$'s corresponding to
$K^+K^-$ pairs are required to be within 30~MeV of the $\phi$ mass.
For the $K^+K^-\pi^0$ channel, at least one of the three $\gamma\gamma$
invariant mass combinations is required to be within 40 MeV of the $\pi^0$ mass;
for events where this happens for more than one combination, 
the one with invariant mass closest to the $\pi^0$ mass is taken to be 
the candidate $\pi^0$.

\begin{figure}[htbp]
  \epsfxsize=3.6in
  \centerline{\epsfbox{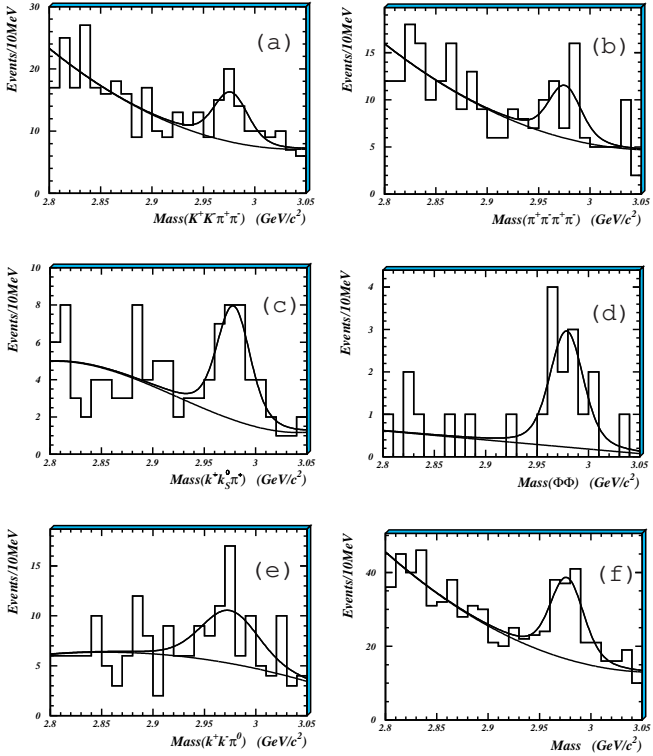}}
  \caption{The (a) $m_{K^+K^-\pi^+\pi^-}$,   (b) $m_{\pi^+\pi^-\pi^+\pi^-}$,
                 (b) $m_{K^\pm K^0_S\pi^\mp}$, (d) $m_{\phi\phi}$ and 
                 (e) $m_{K^+K^-\pi^0}$ distribution in the $\eta_c$ region;
                 (f) is the combined four-charged-track mass distribution
                  of (a), (b), (c) and (d).}
  \label{etac-mass}
\end{figure}

Using the event selection criteria described above, we determine
the invariant mass spectra for each decay mode
shown in Figs.~\ref{etac-mass}(a) to \ref{etac-mass}(e). 
The curve in each figure indicates the result of a likelihood fit using a
Breit-Wigner line shape  convoluted with
a Gaussian mass resolution function for the $\eta_c$, plus a polynomial function to
represent the background.   In these fits, the $\eta_c$ total width 
is fixed at its PDG central value of $\Gamma=13.2$~MeV, and the resolution
at the Monte Carlo determined value. 
The number of fitted events and the  mass of the $\eta_c$ 
determined for each of the 
channels are listed in Table~\ref{Tab2}.  The experimental resolution,
which varies from channel to channel, is also listed in the table.

\begin{table*}[htbp]
\caption{The number of fitted events and the mass for
individual channels.  The errors are statistical.
$\Gamma$ is the full width of the $\eta_c$ fixed at the PDG
value. $\sigma$ is the mass resolution given by the Monte Carlo
simulation.}
\label{Tab2}
\begin{center}
\begin{tabular}{|c|c|c|c|c|}
\hline
Channel & No. of events~~~~~~~~ & mass(MeV)~~~~~~~ & $\Gamma$(MeV)~~~~~~~ & $\sigma$(MeV)~~~~~~~ \\ \hline
$K^+K^-\pi^+\pi^-$                                  & $37.3\pm13.4$~~~~~~~ & $2976.6\pm6.3$~~~~~~~ & 13.2~~~~~~~~ & 13.7~~~~~~~~ \\ \hline
$\pi^+\pi^-\pi^+\pi^-$                              & $24.9\pm11.2$~~~~~~~ & $2975.5\pm7.3$~~~~~~~ & 13.2~~~~~~~~ & 12.8~~~~~~~~ \\ \hline
$K^\pm K_{S}^{0}\pi^\mp \to K^\pm\pi^\mp\pi^+\pi^-$ & $27.5\pm10.4$~~~~~~~ & $2978.6\pm5.2$~~~~~~~ & 13.2~~~~~~~~ & 13.1~~~~~~~~ \\ \hline
$\phi\phi\to K^+K^-K^+K^-$                          & $12.4\pm 3.3$~~~~~~~ & $2978.7\pm6.1$~~~~~~~ & 13.2~~~~~~~~ & 13.2~~~~~~~~ \\ \hline
$K^+K^-\pi^0$                                       & $39.4\pm15.2$~~~~~~~ & $2975.1\pm9.9$~~~~~~~ & 13.2~~~~~~~~ & 25.0~~~~~~~~ \\ \hline
\end{tabular}
\end{center}
\end{table*}

Figure~\ref{etac-mass}(f) shows the combined four-charged-track invariant mass
distributions in the $\eta_c$ mass region
for the $K^+K^-\pi^+\pi^-$, $\pi^+\pi^-\pi^+\pi^-$, 
$K^\pm K_{S}^{0}\pi^\mp$ and $\phi\phi$ channels,
which are the those with similar mass resolution.
Here, a likelihood fit
using a $\Gamma$ fixed at the PDG value and a mass resolution ($\sigma$)
fixed at the  value averaged over the four channels
($\sigma_{avg}=13.3$MeV) 
gives a total of $100.9 \pm 19.8$ $\eta_c$ events and a mass
$m_{\eta_c} = 2976.7\pm 3.4$ MeV with 
a $\chi^2/\mbox{dof}=14.2/20$,  corresponding to a confidence
level of 81.9\%. 
If $\sigma=13.3$ MeV is fixed and the  mass, number of events,
and $\Gamma$ are allowed to float, the resulting mass value and
number of events are
$m_{\eta_c}=2976.7\pm3.0$ MeV and $91.5 \pm 21.2$, respectively.
 
The main systematic errors associated
with the $m_{\eta_c}$ determination arise from the
mass-scale calibration, the detection efficiency,  and the uncertainties
associated with the selection  of the cut values.  
In the case of the $\psi(2S)$ measurement~\cite{bespsi},
the level of the systematic error on the overall mass scale
of BES was
estimated as 0.8 MeV by comparing the masses of the $\chi_{c1}$ and
$\chi_{c2}$ charmonium states, detected in the same decay channels, 
with their PDG values.
These masses have been measured in a number of experiments,
and the reported values have good internal consistency.
The systematic error caused by the detection efficiency was determined
to be 0.7 MeV by using a Monte Carlo simulation.
Systematic errors originating from the cut
conditions are mainly from the confidence-level cuts for the constrained 
kinematic fits  and the photon minimum energy requirement.
For example, when the accepted confidence level probability
is varied between  1\% and 10\%,
the central value of $m_{\eta_c}$ shifts by 0.7 MeV.
When the minimum energy of the photon is changed from 30 MeV to 50 MeV,
the central value of $m_{\eta_c}$ shifts by 0.2 MeV.
The systematic errors associated with the uncertainties
in the experimental mass resolution and the full width of the 
$\eta_c$ are small.
When the experimental mass resolution is varied between the 
extreme values of 11.0 and 15.0 MeV, and the full width
is changed from 10.0 to 16.0 MeV, we find that shifts of
the mass are less than 0.2 MeV.
The total overall systematic error of this measurement is
taken to be 1.3 MeV, the sum in quadrature of all contributions.

Combining the weighted average  with the result for the 
$K^+K^-\pi^0$ decay channel (see Table~\ref{Tab2}), we obtain the result
$m_{\eta_c}=2976.6\pm2.9\pm1.3$ MeV for the five channels. 
Combining this result with that from the BES analysis of 
$\psi(2S)\to\gamma\eta_c$, namely $m_{\eta_c} = 2975.8\pm 3.9 \pm1.2$ MeV
\cite{bespsi}, we obtain
a weighted average $m_{\eta_c} = 2976.3\pm2.3\pm1.2 $ MeV.
Here, since most of the systematic error in the mass scale is
common between the $J/\psi$ and $\psi(2S)$ measurements,
we take the systematic error of the combined measurement to be
that from the $\psi(2S)$
measurement.

\begin{figure}[htbp]
  \epsfxsize=2.75in
  \centerline{\epsfbox{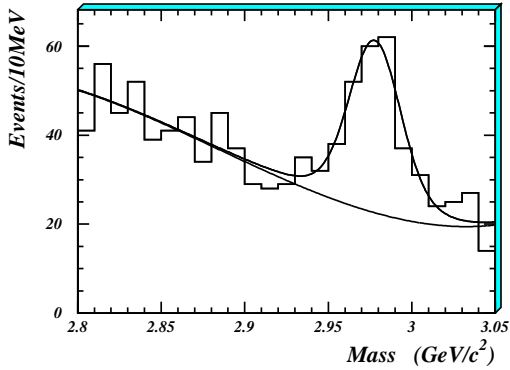}}
  \caption{The combined four-charged-track invariant mass distribution
   in the $\eta_c$ region for $J/\psi\to\gamma\eta_c$ (with
   $\eta_c\to K^+K^-\pi^+\pi^-$, $\pi^+\pi^-\pi^+\pi^-$,
   $K^\pm K_{S}^{0}\pi^\mp$ and $\phi\phi$) and $\psi(2S)\to\gamma\eta_c$
   (with $\eta_c \to K^+K^-\pi^+\pi^-$, $\pi^+\pi^-\pi^+\pi^-$,
   $K^\pm K_{S}^{0}\pi^\mp$ and $K^+K^-K^+K^-$).}
  \label{etac-total}
\end{figure}

The full width of the $\eta_c$ was determined from a fit to
the combined $J/\psi$ and $\psi(2S)$ event samples.
Figure~\ref{etac-total} shows the combined four-charged-track invariant mass
distribution  in the $\eta_c$ mass region for
$J/\psi\to\gamma\eta_c$ (with $\eta_c\to
K^+K^-\pi^+\pi^-$, $\pi^+\pi^-\pi^+\pi^-$,
$K^\pm K_{S}^{0}\pi^\mp$ and $\phi\phi$)
and $\psi(2S)\to\gamma\eta_c$
(with $\eta_c \to K^+K^-\pi^+\pi^-$, $\pi^+\pi^-\pi^+\pi^-$, 
$K^\pm K_{S}^{0}\pi^\mp$ and $K^+K^-K^+K^-$). 
An $\eta_c$ full width of $\Gamma = 11.0\pm 8.1$ MeV is given by a
likelihood fit performed with the resolution fixed at $\sigma = 13.3$~MeV. 
This fit gives a total  of $168.3\pm26.8$ $\eta_c$ events
with a $\chi^2/\mbox{dof}=15.0/21$,
corresponding to a confidence level of 82.1\%.
The systematic error of the width measurement
is 4.1 MeV which includes the sum in quadrature of
the uncertainty in the mass resolution $\sigma$
(2.5 MeV), the uncertainty associated with the choice of selection cuts
(2.5 MeV),
and the mass dependence of the detection efficiency (2.0 MeV).

\begin{figure}[htbp]
\epsfxsize=2.75in
\centerline{\epsfbox{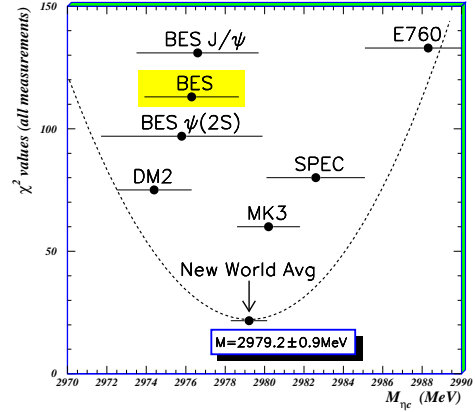}}
\caption{The curve of ${\chi}^2$ versus $m_{{\eta}{c}}$ for a fit that
includes all existing measurements and BES measurements (BES $J/\psi$ 
and BES $\psi(2S)$);
the BES combined results and the four other results with the smallest
errors were indicated as data points. 
(The height of data point has no meaning here.)}
\label{Fig4}
\end{figure}

In summary, we have used the BES 7.8 million $J/\psi$ data sample
to observe the $\eta_c$ in five different 
decay modes and determine the $\eta_c$ mass
to be $2976.6\pm2.9\pm1.3$ MeV. 
Combining this result with
a prior BES analysis of $\psi(2S)\to\gamma\eta_c$, we find
$m_{\eta_c} = 2976.3\pm2.3\pm1.2$ MeV. Combining the two samples, we also obtain
$\Gamma_{\eta_c} = 11.0\pm 8.1\pm 4.1$ MeV. 
The mass measurement of $\eta_c$ from BES is in good agreement with the 
PDG value of $2979.8\pm2.1$ MeV, but 3.8$\sigma$ below the E760 result of
$2988.3_{-3.1}^{+3.3}$ MeV. 
Figure~\ref{Fig4} shows the BES results together with the four previous
measurements
with the smallest errors. The curve in Fig.~\ref{Fig4} allows a determination 
of the values of $\chi^2$ versus $m_{\eta_c}$ for a fit including all existing
measurements. The minimum value, $\chi^2/\mbox{dof}=22.2/8$ occurs at
$2979.2\pm0.9$ MeV. The high $\chi^2$ value is predominantly due to the
poor agreement between the DM2 and E760 measurements. The two
measurements of BES reduce the new world average for the mass by 0.6 MeV. 

   The BES collaboration acknowledges financial support from the Chinese
Academy of Sciences, the National Natural Science Foundation of China, the
U.S. Department of Energy and the Ministry of Science \& Technology of Korea.
It thanks the staff of BEPC for their hard efforts.
This work is supported in part by the National Natural Science Foundation
of China under contracts Nos. 19991480, 19825116 and 19605007
and the Chinese Academy of Sciences under contract No. KJ 95T-03(IHEP);
by the Department of Energy under Contract Nos.
DE-FG03-92ER40701 (Caltech), DE-FG03-93ER40788 (Colorado State University),
DE-AC03-76SF00515 (SLAC), DE-FG03-91ER40679 (UC Irvine),
DE-FG03-94ER40833 (U Hawaii), DE-FG03-95ER40925 (UT Dallas);
and by the Ministry of Science and Technology of Korea under Contract
KISTEP I-03-037(Korea).
We also acknowledge Prof. D.~V. Bugg, Prof. B.~S. Zou and Prof. S.~F. Tuan
for helpful suggestions and discussions.

\end{document}